\documentclass[twocolumn]{aastex62}

\newcommand{\cp}{\citep}
\newcommand{\ct}{\citet}
\usepackage{amsmath}
\usepackage{ mathrsfs }
\usepackage{chngcntr}
\usepackage{appendix}

\graphicspath{{./}{figures/}}

\accepted{May 5, 2020}

\submitjournal{ApJ}

\shorttitle{Bayesian $Kepler$ Occurrence Rates}
\shortauthors{Shabram et al.}

\newcommand{\changes}[1]{#1}

\begin{document}

\title{Sensitivity Analyses of Exoplanet Occurrence Rates from \textit{Kepler} and \textit{Gaia}}

\correspondingauthor{Megan Shabram}
\email{megan.i.shabram@nasa.gov}

\author[0000-0003-1179-3125]{Megan I. Shabram}
\affil{NASA Ames Research Center, Moffett Field, CA 94035, USA}

\author[0000-0002-7030-9519]{Natalie Batalha}
\affil{Department of Astronomy \& Astrophysics, University of California,
Santa Cruz, USA 95064, Natalie.Batalha@ucsc.edu}

\author{Susan E. Thompson}
\affil{Space Telescope Science Institute, 3700 San Martin Drive, Baltimore, MD 21218}

\author[0000-0003-3447-1890]{Danley C. Hsu}
\affil{Department of Astronomy \& Astrophysics, 525 Davey Laboratory, The Pennsylvania State University, University Park, PA, 16802, USA}
\affil{Center for Exoplanets and Habitable Worlds, 525 Davey Laboratory, The Pennsylvania State University, University Park, PA, 16802, USA}
\affil{Center for Astrostatistics, 525 Davey Laboratory, The Pennsylvania State University, University Park, PA, 16802, USA}
\affil{Institute for CyberScience, The Pennsylvania State University}

\author[0000-0001-6545-639X]{Eric B. Ford}
\affil{Department of Astronomy \& Astrophysics, 525 Davey Laboratory, The Pennsylvania State University, University Park, PA, 16802, USA}
\affil{Center for Exoplanets and Habitable Worlds, 525 Davey Laboratory, The Pennsylvania State University, University Park, PA, 16802, USA}
\affil{Center for Astrostatistics, 525 Davey Laboratory, The Pennsylvania State University, University Park, PA, 16802, USA}
\affil{Institute for CyberScience, The Pennsylvania State University}

\author[0000-0002-8035-4778]{Jessie L. Christiansen}
\affil{Caltech/IPAC-NASA Exoplanet Science Institute, 770 S. Wilson Ave, Pasadena, CA 91106, USA} 

\author[0000-0001-8832-4488]{Daniel Huber}
\affil{Institute for Astronomy, University of Hawaii, 2680 Woodlawn Drive, Honolulu, Hawai`i 96822, USA}
\affil{Sydney Institute for Astronomy (SIfA), School of Physics, University of Sydney, NSW 2006, Australia}
\affil{SETI Institute, 189 Bernardo Avenue, Mountain View, CA 94043, USA}
\affil{Stellar Astrophysics Centre, Department of Physics and Astronomy, Aarhus University, Ny Munkegade 120, DK-8000 Aarhus C,
Denmark}

\author[0000-0002-2580-3614]{Travis Berger}
\affil{Institute for Astronomy, University of Hawaii, 2680 Woodlawn Drive, Honolulu, Hawai`i 96822, USA}

\author[0000-0001-7980-4658]{Joseph Catanzarite}
\affil{NASA Ames Research Center, Moffett Field, CA 94035, USA}

\author[0000-0003-3010-2334]{Benjamin E. Nelson}
\affil{Center for Interdisciplinary Exploration and Research in Astrophysics (CIERA), Department of Physics and Astronomy, Northwestern
University, 2145 Sheridan Road, Evanston, IL 60208, USA}

\author[0000-0003-0081-1797]{Steve Bryson}
\affil{NASA Ames Research Center, Moffett Field, CA 94035, USA}

\author{Ruslan Belikov}
\affil{NASA Ames Research Center, Moffett Field, CA 94035, USA}

\author[0000-0002-7754-9486]{Chris Burke}
\affil{MIT Kavli Institute for Astrophysics and Space Research, 77 Massachusetts Avenue, 37-241, Cambridge, MA 02139}

\author[0000-0003-1963-9616]{Doug Caldwell}
\affil{NASA Ames Research Center, Moffett Field, CA 94035, USA}

\begin{abstract}

We infer the number of planets-per-star as a function of orbital period and planet size using \textit{Kepler} archival data products with updated stellar properties from the $Gaia$ Data Release 2.     
Using hierarchical Bayesian modeling and Hamiltonian Monte Carlo, we incorporate planet radius uncertainties into an inhomogeneous Poisson point process model.  
We demonstrate that this model captures the general features of the outcome of the planet \changes{formation and evolution} around GK stars, and provides an infrastructure to use the \textit{Kepler} results to constrain analytic planet distribution models.  
We report an increased mean and variance in the marginal posterior distributions for the number of planets per $GK$ star when including planet radius measurement uncertainties.  
We estimate the number of planets-per-$GK$ star between 0.75 and 2.5 $R_{\oplus}$ and 50 to 300 day orbital periods to have a $68\%$ credible interval of $0.49$ to $0.77$ and a posterior mean of $0.63$. 
This posterior has a smaller mean and a larger variance than the occurrence rate calculated in this work and in \ct{Burke2015} for the same parameter space using the $Q1-Q16$ (previous $Kepler$ planet candidate and stellar catalog).  
\changes{We attribute the smaller mean to many of the instrumental false positives at longer orbital periods being removed from the $DR25$ catalog.} 
We find that the accuracy and precision of our hierarchical Bayesian model posterior distributions are less sensitive to the total number of planets in the sample, and more so on the characteristics of the catalog completeness and reliability and the span of the planet parameter space. 
\end{abstract}


\section{Introduction} \label{sec:intro}

NASA's Kepler Mission was designed to yield an ensemble of planetary systems amenable to statistical analysis \cp{Borucki2010, Koch2010, Jenkins2010a}.  During its primary phase,
$Kepler$ stared nearly continuously at a single field for 4 years, monitoring approximately 190,000 stars that are mostly on the main-sequence \cp{Batalha2010,Brown2011}.
$Kepler$'s goal was to look for signs of transiting exoplanets and ultimately determine the frequency of temperate, Earth-size planets around Sun-like stars. 
This process led to a survey catalog of planet candidates with well-characterized completeness and reliability \cp{Christiansen2017, BurkeJCat2017, Coughlin2017, Mullally2017, Bryson2017}.
Furthermore, \ct{Burke2015} investigated systematics in the derived occurrence rates caused by assumptions about the pipeline sensitivity, characterized by \ct{Christiansen2015}.
The characterization of the $Kepler$ pipeline sensitivity is critical to robust occurrence rate studies, and future work that utilizes the results from $Kepler$. 

With approximately 2327 confirmed planets and 2244 planet candidates from the Kepler Mission \cp{Borucki2011b, Borucki2011a, Batalha2013, Batalha2014, Burke2014a, Rowe2015, Mullally2015, Borucki2016}, scientists are working to incorporate planet \changes{formation and subsequent evolution} theories that can explain both the configuration of our solar system and planetary systems that can be very different from our own. 
For example, systems with dwarf stars and bright giants \cp{Dressing2015, Aguirre2017}, single and binary host stars \cp{Doyle2011, Welsh2012, Orosz2012a, Orosz2012b, Welsh2015}, the number of planets in a system \cp{Lissauer2014, Fabrycky2014}, planet mass and size \cp{Weiss2014, Rogers2015, Wolfgang2016, Carrera2018}, and orbital characteristics \cp{VanEylen2015, Shabram2016}. 
However, large uncertainties in stellar properties translate into large uncertainties in individual planet properties \cp{Huber2014, Berger2018, Fulton2018}, and can limit studies attempting to characterize the exoplanet population.  
Despite the large uncertainties, we are able to develop generative models (i.e., the statistical process that describes how the data are generated) that handle large measurement uncertainty and highly correlated uncertainty of some planet candidate parameters.
Additionally, sources of bias can be naturally incorporated into statistically robust occurrence rate analyses \cp{Youdin2011, ForemanMackey2014, Burke2015, Hsu2018,Hsu2019}.  
These population analyses are becoming more tractable, enabling a better understanding of the physical and orbital properties of exoplanet systems on a broad scale.

Standard occurrence rate studies have largely ignored the radius uncertainty contribution from the planet's host star \cp{Catanzarite2011, Howard2012, Dong2013, Petigura2013a, Petigura2013b,  Farr2014, Dressing2013, Silburt2015, Dressing2015, Mulders2015, Farr2015, Fulton2017, VanEylen2017, Mulders2018, Mulders2019}.
\ct{Burke2015} characterize terrestrial planet occurrence rates for the \textit{Kepler} $GK$ dwarf sample, also without the inclusion of planet radii measurement uncertainties. 
\changes{\ct{Mulders2018} and \ct{Mulders2019} use a forward model with the latest \textit{Kepler} data products to characterize planetary systems around stars (in addition to the number of planets per stellar type). }
\ct{Fulton2018} have investigated the stellar mass dependence of the planet radius gap using $Gaia$ updated stellar mass, stellar radius and planet sizes for the \textit{Kepler} sample.
However, \ct{Fulton2018} do not include the impact of planet radius uncertainties, accounting for survey completeness in an inverse detection efficiency method, a method shown to bias occurrence rates towards smaller values in \ct{ForemanMackey2014, Hsu2018}. 

The $Gaia$ data release 2 has now provided more precise stellar measurement uncertainties \cp{GaiaDR2, Berger2018}. 
Updates to the stellar properties in the \textit{Kepler} sample now enable more robust hierarchical Bayesian occurrence rate posterior distributions. 
The contribution to occurrence rate estimates from uncertainty in planet radius can be included in occurrence rate estimates by using the uncertainty in the measured planet-to-star radius ratio from transit light curve modeling.
To get the planet radius, the planet-to-star radius ratio is simply multiplied by the assumed host star radius point estimate. 
This has been done in \ct{Hsu2018}, an approximate Bayesian computation occurrence rate analysis for $GK$ stars.
\ct{ForemanMackey2014} consider the contribution to the planet radius uncertainties from the measured planet-to-star radius ratio and stellar radius uncertainties in their occurrence rate analysis for $GK$ stars. 
However, they use a non-parametric Bayesian method that makes it difficult to interpret some population level parameters for planet \changes{formation and subsequent evolution} theories. 
\ct{Hsu2019} use approximate Bayesian computation to include the host star radius uncertainties and planet-to-star radius ratio uncertainties by incorporating additional \textit{Kepler} data products to accurately characterize the the efficiency of planets being recognized as a `threshold crossing events' (TCE).
\changes{In addition, \ct{Hsu2019} include catalog reliability in the occurrence rate calculations, which is also another important parameter that can affect the results.}

Including measurement uncertainties in the occurrence rate calculations is impactful for many reasons. 
When using the $Kepler$ catalog of planet candidates to constrain hierarchical Bayesian models, we are able to marginalize over noise when reporting posteriors of the number of planets-per-star. 
Including the measurement uncertainty is necessary to avoid a bias due to only using a histogram of mean values to infer population distributions.  
Furthermore, the inclusion of measurement uncertainties can allow better exploration of population level parameters that describe planet \changes{formation and subsequent evolution} relations.  

In this work, we use Hamiltonian Monte Carlo (HMC) \citep{Neal2012, StanPaper} to perform hierarchical Bayesian model calculations.   
The Hamiltonian Monte Carlo method is the state-of-the-art for sampling hierarchical Bayesian models. 
HMC uses a kinetic energy term, taking advantage of the gradient of the target density to efficiently sample from high dimensional posteriors. 
For example, HMC can handle the inclusion of measurement uncertainties and many population-level parameters, for likelihood-based continuous distribution models.  
Furthermore, HMC provides advanced diagnostics to look for sources of numerical bias and other model pathologies characteristic to using MCMC methods to perform hierarchical Bayesian model calculations.   
Thus, Hamiltonian Monte Carlo is a powerful sampling method and very applicable for this work. 

Here, we employ a hierarchical Bayesian model in conjunction with a Hamiltonian Monte Carlo sampler to infer planet occurrence rates while including the contribution from the planet host star radius uncertainty into the uncertainties in planet size.
We demonstrate the use of standard and advanced diagnostics to assess the application of Hamiltonian Monte Carlo for performing our hierarchical Bayesian model calculations.  
We use this statistical framework to demonstrate the impact of subtle differences in host star categorization and small differences in selected planet radii and orbital period across varied completeness and reliability parameter spaces. 

In \S \ref{sec:methods}, we describe the observations and parameter space used in our investigations. 
In \S \ref{sec:statsframe} we explain the statistical framework for this work.  
In \S \ref{sec:results} we explore the sensitivity of our occurrence rate methodology to small changes in the selected stars, reliability and completeness, the number of planets, and uncertainties in planets size.  
In \S \ref{sec:discussion} we discuss our experimental design and future research. 
In \S \ref{sec:conclusions} we summarize the conclusions of this work.

\section{Observations} \label{sec:methods}

In \S \ref{sec:stars} through \S \ref{obs:detmodel}, we describe the various stellar cuts, planet parameter cuts, and the detection model used in this work.
We use the cuts described below to explore the sensitivity of posterior estimates of occurrence rates from our statistical framework to subtle changes in the selected stars, selected planet parameters, the inclusion of radius measurement uncertainties, and updated stellar properties from $Gaia$. 

\subsection{Stars} \label{sec:stars}

We apply our model to three stellar catalogs with two sets of stellar cuts. 
A summary of the stellar cuts can be found in Table \ref{table1:StarRanges} and a summary of the catalogs used can be found in Table \ref{table2:StarCutNumbers}.
The first set of stellar cuts (labeled ``$GK \;cuts\, \uparrow$'') describes stellar cuts similar to those used in \ct{Burke2015} and \ct{Hsu2018} using the $Q1-Q16$ catalog release \cp{Mullally2016}.
The up arrow indicates that this selection of $GK$ stars has more stars compared to our second definition of $GK$ stars, which we label ``$GK \;cuts\, \downarrow$''. 
This second case contains less stars, and is similar to the cuts used in the NASA Exoplanet Program's Study Analysis Group 13\footnote{\href{https://exoplanets.nasa.gov/system/internal\_resources/details/original/680\_SAG13\_closeout\_8.3.17.pdf}{https://exoplanets.nasa.gov/system/ \\ internal\_resources/details/original/680\_SAG13\_closeout\_8.3.17.pdf}} (see Table \ref{table2:StarCutNumbers}). 
We choose these two selections to investigate how sensitive our results are to relatively small differences in the definition of the stellar category of interest, and to explore how much power the data has to explore trends in stellar properties while using the state-of-the-art \textit{Kepler} planet and star catalogs.

\begin{deluxetable}{cc}
\tablecaption{Summary of GK Star Classifications}
\tablenum{1}

\tablehead{ \colhead{$stars\;(GK \;cuts)\uparrow$ } & \colhead{$stars\;(GK \;cuts)\downarrow$}   } 

\startdata
$T_{eff}: 4200 - 6100 K$  &   $T_{eff}: 3900 - 6000 K$ \\
$R_{*}$ \textless $\;1.15$ & $R_{*}$ \textless $\;1.35$ \\
$log\;g\;$\textgreater $\;4.0$ &   $log\;g\;$\textgreater $\;3.8$ \\
\enddata


\tablecomments{``$GK \;cuts\, \uparrow$'' are similar to the stellar parameter cuts used in the occurrence rate studies for the $Q1-Q16$ $Kepler$ planet candidate catalog release \cp{Mullally2016}. ``$GK \;cuts\, \downarrow$'' are similar to the stellar parameter selection used in the $SAG\, 13$ analysis to compare occurrence rates across different teams. }

\label{table1:StarRanges}
\end{deluxetable}

Before selecting the $GK$ stars to be analyzed with the updated $Gaia$ stellar properties, we start with a selection of $FGK$ stars from the \textit{Gaia Data Release 2} \cp{GaiaDR2doc, GaiaDR2} cross-matched to the $Kepler$ $DR25$ stellar catalogs \cp{Mathur2017}. 
\changes{All stars included in these cross-matched catalogs use the available $Gaia$ stellar properties.} 
Initially, the cross-match between the $Kepler$ and $Gaia$ catalogs is based on position alone.  
For some $Kepler$ targets, there are multiple $Gaia$ targets that match positionally.  
To uniquely identify the source, we computed a delta magnitude and looked at its distribution.  
We use various quality cuts that further reduce our crossmatched sample.
The motivation for these cuts is to choose a sample of stars where we are reasonably confident that each is near the $FGK$ main sequence and is less likely to be impacted by sources of dilution.  Both a maximum parallax uncertainty ($10\%$) and the GAIA data quality flags are chosen so as to provide a cleaner sample.  
For instance, binary stellar companions can contribute to excess scatter about the astrometric model.  
\changes{We note that unlike Berger et al. (2018), no extinction corrections were applied. }
This results in a set of 78,005 $Kepler$ target stars\footnote{A table listing the 78,005 targets with their KIC and Gaia ID's, parameters and parameter uncertainties can be found at: \href{https://github.com/mshabram/PyStan\_Kepler\_Exoplanet\_Populations/blob/master/Sensitivity-Analyses-of-Exoplanet-Occurrence-Rates-from-Kepler-and-Gaia/Data/q1q17\_dr25\_gaia\_fgk.csv}{github.com/mshabram/PyStan\_Kepler\_ \\ Exoplanet\_Populations/blob/master/Sensitivity-Analyses-of-Exoplanet-Occurrence-Rates-from-Kepler-and-Gaia/Data/q1q17\_dr25\_gaia\_fgk.csv}.}. 
These selection criteria are applied in the following order:

\begin{itemize} 
\item First, we remove all duplicate $Gaia$ source ID rows (these duplicates also share $Kepler$ IDs).
\item We make a cut where the difference for all crossmatched targets between the $Gaia$ $G$ mean magnitude and the $Kepler$ magnitude (with bandpasses that have similar overall shape, range, and median) is within 1.5-sigma of the median.
We chose this threshold for ($Gaia$ G)-($Kepler$ Mag) that prevents matching more than one $Gaia$ target to our $Kepler$ targets, thus preventing us from using stellar properties associated with a background or foreground star rather than the intended $Kepler$ target.  
We address the slight difference between the $Gaia$ $G$ and $Kepler$ by using the median of the differences. 
\item Following \ct{Evans2018} we select on Astrometric Goodness of Fit in the Along-Scan direction ($GOF\_\,AL$) of less than 20, and on Astrometric Excess Noise of less than 5, to exclude potential poorly-resolved binaries or other problematic targets.
\item We include parallax quality cuts using the processing flag outputs of the module that calculates astrophysical parameters for the $Gaia$ target stars. 
We selected only targets for which the Priam processing flags (A and B) are zero.
This selects strictly positive parallax values, colors close to the standard locus, and parallax error less than 0.05 mas \cp{Lindegren2018, Andrae2018}.
We note that the sky position of the target stars does not change much over the full $Kepler$ field.
We assume that occurrence rates don't depend on a star's position in the galaxy, so the dependence of parallax error on sky position does not introduce significant bias.  
This would become important for assessing occurrence rates between disk and halo stars. 
\item Sources with $Kepler$ magnitude less than 16 are removed, and we apply a magnitude cut of $0.5 < G_{bp}-G_{rp} < 1.7$ \cp{Lindegren2018}.
This color cut is more precise than using the temperature from the Kepler Input Catalog and more uniform than using temperatures from the DR25 stellar catalog, for selecting $FGK$ stars.
\item Furthermore, we use a six iteration quadratic fit of the color-luminosity relation for the main sequence with $log_{10}(1.75)$ width to select $FGK$ targets.
\end{itemize}

We summarize the stellar catalog versions investigated in this work in Table \ref{table2:StarCutNumbers}, and report the number of selected stars for each case. 
Here, ``$Q1-Q16$'' refers to the version of the \textit{Kepler} star and planet catalogs release that precedes the ``$DR25$'' catalog release. 
We evaluate occurrence rates for the $DR25$ and ``$DR25+Gaia$'' (a version that uses \textit{Gaia} updated stellar properties) catalogs with the ``$GK \;cuts\, \downarrow$'' selections that were designated during the The NASA Exoplanet Exploration Program Analysis Group ($ExoPAG$) Study Analysis Group 13 ($SAG\, 13$) working group meeting. 
In this work, we analyze the $Q1-Q16$ planet candidate catalogs to benchmark our methods and results against the previous work of \ct{Burke2015} and \ct{Hsu2018}.
Therefore, we only consider the ``$GK \;cuts\, \uparrow$'' case ($T_{eff}: 4200 - 6100 K$, $R_{*}$ \textless $\;1.15$, and $log\;g\;$\textgreater $\;4.0$) with the  $Q1-Q16$ planet and star catalog.
By comparing the $Q1-Q16$ planet candidate catalog occurrence rates to occurrence rates using the $DR25$ planet candidate catalog, we can see the impact on occurrence rates when many of the instrumental false positives at longer orbital periods have been removed from the $DR25$ catalog.
The vetting process and reliability characterization can be found in \ct{Thompson2018}. 

In \S 4.6 we compare occurrence rate posteriors using the catalogs described here to the catalog provided in \ct{Berger2018}.  
The \ct{Berger2018} catalog updates host star radius values using values that were spectroscopically derived in the \textit{California-Kepler Survey} (CKS) \cp{Fulton2017, Petigura2017, Johnson2017} \changes{as well as stars cooler than $4,000 K$ from \ct{Gaidos2016}. }
However, the full population of stars searched by \textit{Kepler} has not been updated with spectroscopic followup at this time.

\begin{deluxetable*}{lcc}
\tablecaption{Summary of Selected Stars from Various Stellar Catalogs}
\tablenum{2}

\tablehead{\colhead{Catalog} & \colhead{\#\,$stars\;(GK \;cuts)\uparrow$ } & \colhead{\#\,$stars\;(GK \;cuts)\downarrow$}   } 

\startdata
a. $Q1-Q16$ & $91,446$ & $N/A$  \\
b. $DR25$ & $88,807$ & $81,882$  \\
c. $DR25+Gaia$ & $N/A$ & $44,597$ \\
\enddata


\tablecomments{We can compare results across disparate stellar catalogs using hierarchical Bayesian analysis.}

\label{table2:StarCutNumbers}
\end{deluxetable*}

\subsection{Planets}  \label{sec:planets}

We choose two different cuts in planet parameters.  
First, we consider planets with sizes that range from $1$ to $2$ $R_{\oplus}$ and orbital periods that range from $50$ to $200$ days, referred to as the ``$planets\downarrow$'' case in Table \ref{table3:PlanetRanges} and here after.   
These cuts span a parameter space for $GK$ stars that has a slightly higher average detection completeness than the second case we investigate.
The second case we refer to as ``$planets\uparrow$'', which includes planets with sizes between $0.75$ to $2.5$ $R_{\oplus}$ and orbital periods between $50$ to $300$ days. 
This case now contains less reliable planet candidates and has a larger variance in completeness values across the planet parameter space. 
In this case, the top left corner of the completeness grid \cp{Thompson2018} near $P_{orb}$ = 50 days and $R_{p}$ = 2.5 $R_{\oplus}$ has a higher reliability and completeness while the opposite corner near $P_{orb}$ = 300 days and at $R_{p}$ = 0.75 $R_{\oplus}$ has a lower reliability and completeness. 
The planets ``$planets\downarrow$'' case is contained within the ``$planets\uparrow$'' case and has overall less variance than the ``$planets\uparrow$'' case. 
The detection completeness model is discussed further in \S \ref{obs:detmodel}.
These cuts were chosen to compare to previous work and to assess how subtle differences in the completeness and reliability and in the ranges in planet parameter space can influence occurrence rate posteriors.

\begin{deluxetable*}{ccc}
\tablecaption{Summary of Planet Size and Orbital Period Ranges}
\tablenum{3}

\tablehead{ \colhead{} & \colhead{$R_{p_{\;min}} - R_{p_{\;max}} [R_{\oplus}]$}  & \colhead{$P_{orb_{\;min}} - P_{orb_{\;max}} [Days]$}   } 

\startdata
$planets\downarrow$ & $1.00-2.00 $ & $50-200$ \\
$planets\uparrow$  &  $0.75-2.50 $ & $50-300$ \\
\enddata


\tablecomments{We select these fairly complete orbital period and planet size ranges to facilitate comparison between catalogs and previous work, and assess the sensitivity of occurrence rate posteriors to the choice of planet parameters when using our parametric hierarchical Bayesian model.}

\label{table3:PlanetRanges}
\end{deluxetable*}

\begin{deluxetable*}{lcr}
\tablecaption{Summary of Selected Stars and Planets}
\tablenum{4}

\tablehead{\colhead{Catalog} & \colhead{\#\, $planets\downarrow/stars$ } & \colhead{\#\,$planets\uparrow/stars$ }  } 

\startdata
a. $Q1-Q16 \;(GK \;cuts)\;stars\uparrow$ & $N/A$ & $106 / 91,446 \;(0.0012)$ \\
b. $DR25 \;(GK \;cuts)\;stars\uparrow$ & $54 / 88,807\;(0.0006)$ & $118 / 88,807\;(0.0013)$ \\
b. $DR25 \;(GK \;cuts)\;stars\downarrow$ & $58 / 81,882\;(0.0007)$ & $124 / 81,882\;(0.0015)$ \\
c. $DR25 + Gaia \;(GK \;cuts)\;stars\downarrow$ & $N/A$ & $85 / 44,597\;(0.0019)$ \\
\enddata


\tablecomments{Subtle changes in stellar parameter selections can result in datasets with less stars having more planets. This effect is seen in occurrence rate posteriors suggesting that our method may be sensitive to probing relations with stellar parameters, even when using simple planet \changes{formation and subsequent evolution} distribution models. The differences in the number of selected stars has a negligible contribution to the expected number of planets-per-star due to the contribution from the completeness model used in the hierarchical Bayesian statistical framework. However, subtle differences in the number of selected planets could in part be due to unaccounted for reliability.}

\label{table4:StarPlanetCutNumbers}
\end{deluxetable*}

\subsection{Detection Model} \label{obs:detmodel}

We employ the analytic pipeline completeness model described in \S 2 of \ct{Burke2015} to compare our results against previous catalogs and for sensitivity analysis.
We precompute the completeness over a $61\times57$ (planet radius $\times$ orbital period) grid. 
We approximate the completeness as constant within each bin using the value calculated for each bin center after dividing the planet radius range by 61 and the orbital period range by 57. 
For the gamma cumulative distribution function (CDF) coefficients (shape $a$, $scale$, and $size$) that describes the average detection efficiency of selected $GK$ stars for our $DR25$ and $DR25 + Gaia$ catalog analysis, we use $a=30.87$, $size=0$, and $scale=0.271$, with a plateau factor of $0.94$ \cp{Thompson2018, Christiansen2017}.
These coefficients are derived using a gamma CDF that is fit to a detection efficiency model that includes vetting completeness. 
For our $Q1-Q16$ analysis, we use $a=4.65$, $size=0$, and $scale=0.98$ \cp{Burke2015}. 
We calculate transit durations assuming a circular orbit, and use the mean stellar radii estimates.  
Figure 2 of \ct{Burke2015} shows the absolute difference between the analytic model used in this study, and the higher fidelity completeness model available as part of the DR25 occurrence rate data product release.
Since differences are largest (a relative fraction of approximately 0.06) towards longer orbital periods, we focus our analysis on the parameter space of $P_{orb}\textless300$ days for $GK$ stars.
This allows us to investigate a region of parameter space with relatively high reliability and completeness.  
We have not included a model for reliability in our analyses, however, we have restricted our analyses to shorter orbital periods where reliability is higher based on estimates from \ct{Thompson2018}.
Furthermore, preliminary results show that occurrence rate posteriors are not significantly influenced by reliability when planet orbital periods are less than 300 days. 
A discussion of the impact of the latest DR25 pipeline completeness and reliability products will be available in C. J. Burke et al. (2019, in preparation).
Future studies will explore more vigorous treatments of including vetting efficiency and numerical pipeline completeness models.  
We discuss this further in \S \ref{sec:ongoingwork}.

\section{Statistical Framework} \label{sec:statsframe}

We calculate occurrence rates using the inhomogeneous Poisson point process method with a parametric rate intensity as implemented in \ct{Burke2015, Youdin2011, Gregory1992}, now using Hamiltonian Monte Carlo \citep{Neal2012, StanPaper} and including planet radius measurement uncertainties\footnote{Code can be found at: \\ \href{https://github.com/mshabram/PyStan_Kepler_Exoplanet_Populations}{github.com/mshabram/PyStan\_Kepler\_Exoplanet\_Populations}.}.

\subsection{The Hierarchical Bayesian Model} \label{sec:likelihood}

For this study, we parameterize the rate intensity function of an inhomogeneous Poisson point process as a power law scaling of the planet radius and the orbital period.  
The inhomogeneous Poisson point process is a natural choice of the likelihood function for the occurrence of exoplanets per star, where each planet occurrence that is counted is very nearly independent of each other planet occurrence that is counted (ignoring multiple planet systems).

The likelihood for our model is adopted from \ct{Burke2015} and \ct{Youdin2011}, now with the addition of Gaussian noise in planet size:
\begin{equation}
\mathcal{L} = \Bigg[ \prod_{l=1}^{N_{l}}f_{l}\Bigg]exp(-N_{exp}),
\end{equation}
where $N_{l}$ is the number of selected planets after the cuts in stellar parameters, planet radius, and orbital period have been applied.  
$N_{exp}$ is the number of expected detections in all bins, defined as:
\begin{equation}
\begin{split}
N_{exp} = F_{0}\int\displaylimits_{P_{min}}^{P_{max}} \int\displaylimits_{R_{min}}^{R_{max}} \Bigg[\sum_{j=1}^{N_{*}}n_{j}(P_{orb},R_{p})\Bigg] \\ \times {\Bigg(\frac{P_{orb}}{P_{0}}\Bigg)}^{\beta}{\Bigg(\frac{R_{p}}{R_{0}}\Bigg)}^{\alpha}dP_{orb}dR_{p}
\end{split}
\end{equation}
where $n_{j}$ is the survey completeness (see \S \ref{obs:detmodel} and references therein), which is a function of the planet radius $R_{p}$, orbital period $P_{orb}$, and stellar properties. 
The survey completeness is precomputed outside of our hierarchical Bayesian model as was done in \ct{Burke2015}, and also depends on the stellar mass, stellar radius, and semi major axis.
The hyperparameters in this hierarchical Bayesian model are $\alpha$ (the power law index for the planet radius distribution), $\beta$ (the power law index for the orbital period distribution), and $F_{0}$ (the integrated number of planets per star).
$f_{l}$ is the number density from the power-law scaling of the planet occurrence rate evaluated over the list of detected planets.
As we numerically simulate this likelihood function using Hamiltonian Monte Carlo (see \S \ref{sec:hmc}), each planet can take on values normally distributed around the true planet size $R_{p^{(l)}}$ (with $N_{l}$ latent variables corresponding to the number of planets in the sample) and reported standard deviation $\sigma_{R_{p^{(l)}}}$ of the observed planet radius $R_{P\;obs^{(l)}}$:
\begin{equation}\label{eq:Robs}
\begin{split}
R_{P\;obs^{(l)}} \sim Normal(R_{p^{(l)}}, \sigma_{R_{p^{(l)}}})\;\;: \\ T[R_{p_{\;min}}, R_{p_{\;max}}]
\end{split}
\end{equation}
The convolution of the true planet sizes with their measurement uncertainties are truncated, $T[R_{p_{\;min}}, R_{p_{\;max}}]$, so that draws that are outside the selected planet range (either $planets\uparrow$ or $planets\downarrow$) are not considered when numerically simulating the integral of the likelihood.  
The truncation allows the data to be described as resulting from a data generating process that only produces values within an interval.
In this case values that are drawn below and or above the specified interval are treated as not observed. 
This allows us to investigate how the choices in cuts impact the resultant occurrence rate distributions.  
Creating a model that allows for planets with mean radius values outside the selected parameter space to enter into the calculation of the posterior distribution for the selected range is beyond the scope of this paper.  
In this work, by definition, if the planet?s true value exists inside the selected range, it does not exists outside the selected range.

This hierarchical Bayesian model also ignores the constant multiplicative factors resulting in the survey completeness only entering the equations in the number of expected detections for all bins \cp{Youdin2011}.
We note that the uncertainties in planet size mean that when using our hierarchical Bayesian analysis, there is a non-zero amount of planets that have a non-zero chance of occurring outside of the selected range while sampling from our likelihood function. 
This effect will need to be explored in the future by allowing the number of planets in the sample $N_{l}$ to have flexibility, and to exclude planets where draws do not land inside the given range.
Currently, our method assumes that all the planets selected have true values within the planet radius ranges specified. 
We reason that this effect would be important when stitching together occurrence rate analysis for different planet radius ranges with our current parametric method.

\subsection{Hamiltonian Monte Carlo} \label{sec:hmc}

We use the Stan Bayesian statistical modeling software \cp{StanPaper} to perform numerical calculations.
We utilize the extensive Stan diagnostics to assess of the convergence of our HMC simulations.  
We use uniform priors ranging from $-5$ to $5$ for our hyperparameters $\alpha$, $\beta$ and $ln\,F_{0}$. 
We advance 4 chains for 1500 warm-up iterations followed by 1500 sampling iterations.

The treedepth is a configuration parameter of the No-U-Turn-Sampler used by Stan that can impact efficiency\footnote{A brief guide to Stan's warnings can be found at http://mc-stan.org/misc/warnings.html}.
We set the maximum tree depth to 10.
We increase the maximum to 11, which roughly doubles the compute time.   
Each chain has an energy Bayesian fraction of missing information (E-BFMI) of approximately $0.8$.  
A low E-BFMI ($<0.02$) for a given chain implies a problem with the adaptation phase, and those chains likely did not explore the posterior distribution efficiently \cp{Betancourt2016}.

We obtain Gelman-Rubin statistics $\hat{R}$ of $1.0$ for all parameters, and zero divergent transitions.
Gelman-Rubin statistics are used to evaluate the variance within and between Markov chains. 
Large Gelman-Rubin statistics indicate possible non-convergence.
A Gelman-Rubin value close to 1 indicates no sign of non-convergence from this particular statistical test.
Divergent transitions are an indication that your posterior estimates are biased from numerical error. 
We obtain effective samples sizes (ESS) of approximately 4,600 for $\alpha$, 6,000 for $\beta$, approximately 4,300 for $ln\,F_{0}$, and 6,000 for all the latent variables $R_{p^{(l)}}$. 
The ESS is a measure of how many draws from the Markov chain are effectively independent after the burn-in phase.

\section{Results}  \label{sec:results}

\begin{figure*}
\centering
\begin{minipage}{0.9\textwidth}
\plotone{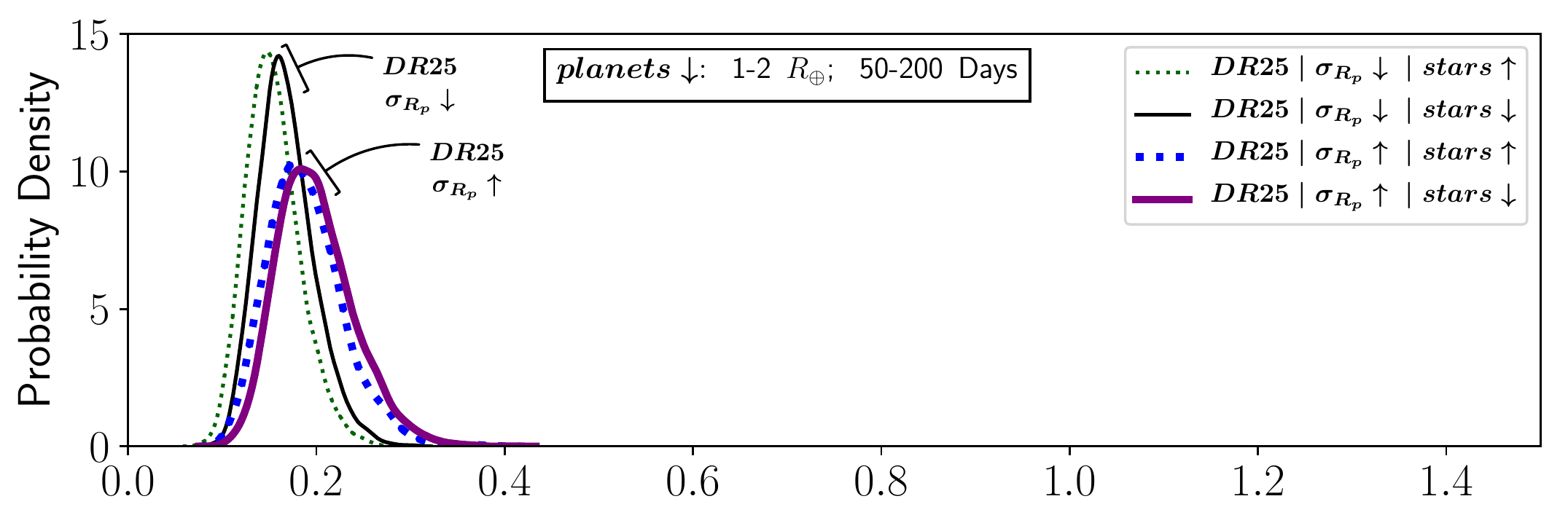}
\end{minipage}
\begin{minipage}{0.9\textwidth}
\plotone{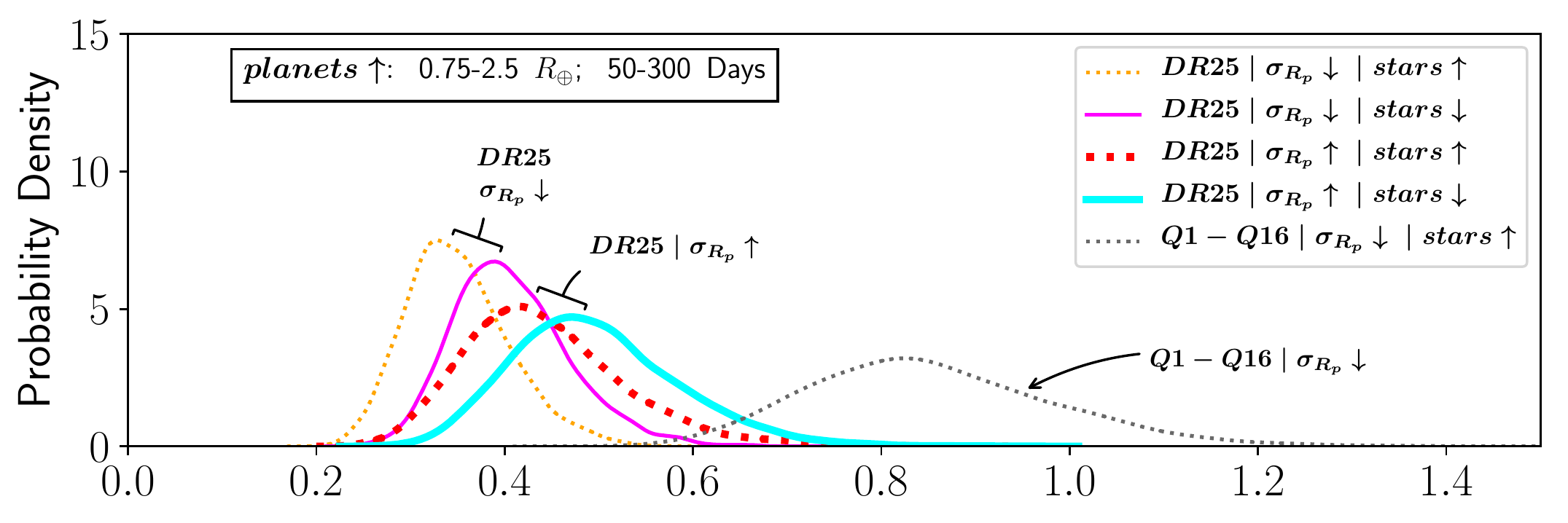}
\end{minipage}
\begin{minipage}{0.9\textwidth}
\plotone{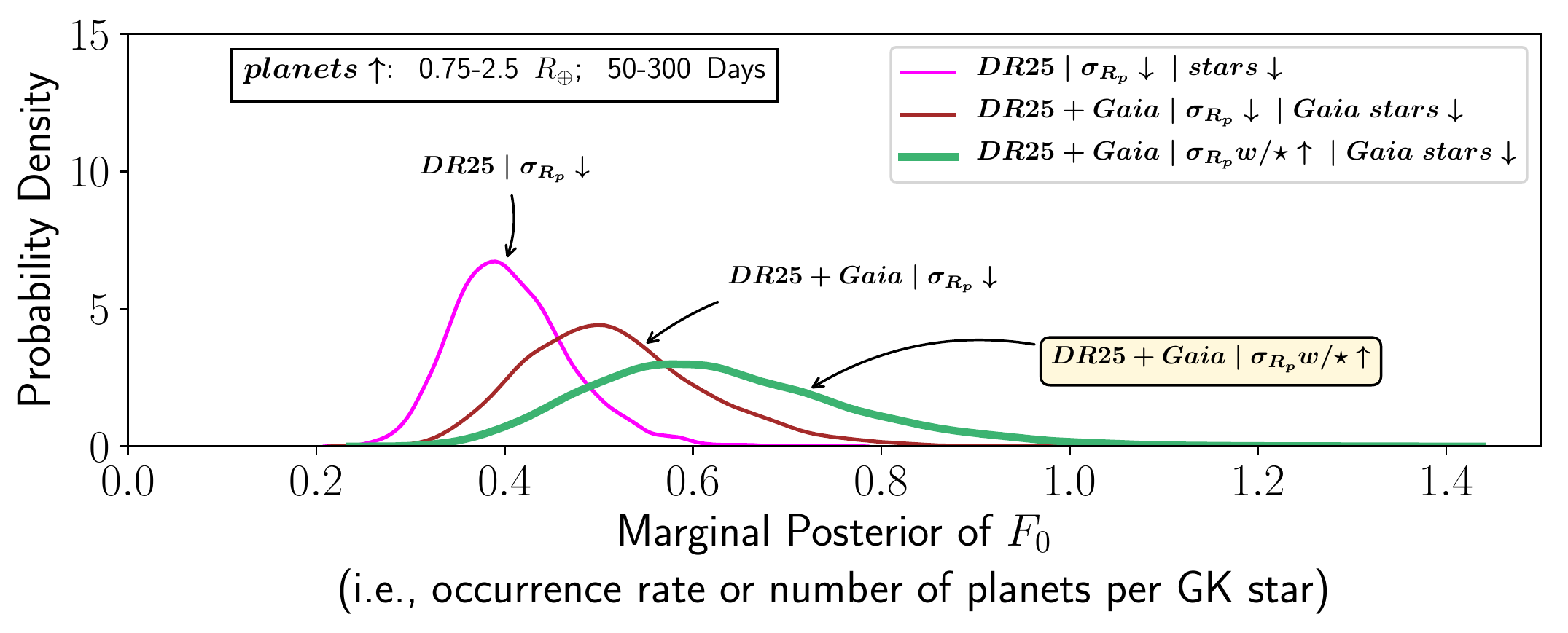}
\end{minipage}
\caption{\textbf{Kernel density estimates of marginal posterior distributions for the number of planets per GK star.} 
The key labels read from top to bottom corresponding to curves going from the left to right. 
$stars\uparrow$ ($stars\downarrow$) means more (fewer) stars.
$planets\uparrow$ ($planets\downarrow$) means more (fewer) planets.
$\sigma\uparrow$ ($\sigma\downarrow$) means with (without) measurement uncertainty in planet size. 
The dashed lines help indicate the occurrence rates calculated using the slightly warmer set of stars, $GK \;cuts\; \uparrow$ (i.e., $stars\uparrow$), described in Table \ref{table1:StarRanges}. 
The thicker lines help indicate the inclusion of planet radius measurement uncertainties (i.e., $\sigma\uparrow$). 
Excluding planet size ($R_{p}$) measurement uncertainty biases occurrence rates towards smaller values: compare dashed-green/dashed-blue and black/purple pairs in the top panel, and dashed-orange/dashed-red and pink/cyan curve pairs in the middle panel. 
These correspond to fixed planet and star cuts with no measurement uncertainty/with measurement uncertainty ($\sigma_{R_{p}}\downarrow$\,/\,$\sigma_{R_{p}}\uparrow$), respectively. 
A previous lower reliability $Kepler$ planet candidate catalog ($Q1-Q16$ catalog) included more false positives, inflating the occurrence rate for this parameter space (dashed-grey curve in middle panel). 
Subtle differences in stellar cuts can impact the number of planets selected, where more stars result in less planets (compare dashed-green/black and dashed-blue/purple curves in the top panel, and dashed-orange/pink and dashed-red/cyan curves in the middle panel). 
The occurrence rate variance is lower for planets in a slightly more complete part of parameter space ($planets\downarrow$ in top panel) than in a slightly less complete part of parameter space ($planets\uparrow$ in middle and bottom panels), even when there are less planets present in the $planets\downarrow$ case. 
Although we expect the ``$planets\uparrow$'' cases to have larger occurrence rates than the $planets\downarrow$ cases (because we are probing a larger domain) the ``$planets\uparrow$'' occurrence rate posteriors could be over estimated due to the low, unaccounted for, reliability in the corner near 0.75 $R_{\oplus}$ and 300 days.
Comparing the pink and brown curves in the bottom panel shows the impact on occurrence rate posteriors in the $planets\uparrow$ and  $stars\downarrow$ parameter space when using updated stellar radii from $Gaia$ in the completeness model and updating planet sizes (and excluding measurement uncertainty). 
Propagating the stellar uncertainties from $Gaia$ into the planet size ($R_{p}$) uncertainties while simultaneously updating stellar radii in the completeness model removes the bias towards smaller values and increases the variance of the occurrence rate (light-green curve in bottom panel). \label{fig:f1}}
\end{figure*}

In order to investigate the sensitivity of occurrence rate posteriors from \textit{Kepler} data to small changes in the selected stars, reliability and completeness, the number of planets, and uncertainties in planets size, we perform fits over both the $R_{p}$ and $P_{orb}$ ranges (``$planets\uparrow$'' and ``$planets\downarrow$'', described in Table \ref{table3:PlanetRanges}).
The posterior distribution for all model parameters including the power-law parameter estimates that describe the general features of the outcome of planet \changes{formation and subsequent evolution} can be found on the github repo for this project\footnote{\href{https://github.com/mshabram/PyStan\_Kepler\_Exoplanet\_Populations/tree/master/Sensitivity-Analyses-of-Exoplanet-Occurrence-Rates-from-Kepler-and-Gaia/posterior-distributions}{github.com/mshabram/PyStan\_Kepler\_Exoplanet\_Populations/ \\ tree/master/Sensitivity-Analyses-of-Exoplanet-Occurrence-Rates-from-Kepler-and-Gaia/posterior-distributions}}.
We assess the occurrence rate posteriors when making subtle changes to the definition of $GK$ stars (described in Table \ref{table1:StarRanges}).  
Figure \ref{fig:f1} shows kernel density estimates of marginal posterior distributions for the occurrence rate (i.e., the number of planets per GK star, $F_{0}$).
The key labels read from top to bottom corresponding to curves going from the left to right. 
The $stars\uparrow$ ($stars\downarrow$) label means more (fewer) stars, the $planets\uparrow$ ($planets\downarrow$) means more (fewer) planets, and the $\sigma\uparrow$ ($\sigma\downarrow$) means with (without) measurement uncertainty in planet size. 
The dashed lines help indicate the occurrence rates calculated using the slightly warmer set of stars, $GK \;cuts\; \uparrow$ (i.e., $stars\uparrow$), described in Table \ref{table1:StarRanges}. 
The thicker lines help indicate the inclusion of planet radius measurement uncertainties (i.e., $\sigma\uparrow$). 
We will refer to this figure in \S \ref{sec:results:senstostars} through \S \ref{sec:results:starsGaia}.
Summary Statistics for the occurrence rate posterior distributions can be found in Table \ref{table5:SumStats} and two sample Kolmogorov-Smirnov (K-S) statistics for pairs of these occurrence rate posterior distributions can be found in Appendix A.

\subsection{Sensitivity to selections in planet radius and orbital period} \label{sec:results:senstoplanetradii}
 
We investigate the sensitivity of occurrence rate posteriors to the range of planet radii and orbital periods by comparing across the two ranges in planet radius and orbital periods described in Table \ref{table3:PlanetRanges}. 
Our ``$planets\downarrow$'' case contains approximately half the number of selected planets as our ``$planets\uparrow$'' case, and lies in a slightly higher average completeness space for $GK$ stars of interest.  
The top panel of Figure \ref{fig:f1} shows the marginal posterior for the number of planets per $GK$ star for which the selected number of planets follows from the ``$planets\downarrow$'' case. 
The middle and bottom panels of Figure \ref{fig:f1} shows occurrence rate posteriors for the number of planets per $GK$ star when using the ``$planets\uparrow$'' case (that includes the planets from the ``$planets\downarrow$'' case). 

When comparing these two clusters of marginal posteriors, we see that the ``$planets\downarrow$'' curves have less variance than the marginal posteriors for the cluster of the ``$planets\uparrow$'' cases, even though the ``$planets\downarrow$'' case contains approximately half the number of selected planets.  
This could be in part due to the completeness and reliability varying more across the ``$planets\uparrow$'' case (the larger planet parameter space box). 
For example, when comparing the completeness between the ``$planets\uparrow$'' case and the ``$planets\downarrow$'' case, parts of the larger box (``$planets\uparrow$'') are in a more complete and higher reliability space (i.e., at $P_{orb}$ = 50 days and at $R_{p}$ = 2.5 $R_{\oplus}$) while another section is in a lower reliability and lower completeness space (i.e., at $P_{orb}$ = 300 days and at $R_{p}$ = 0.75 $R_{\oplus}$).  
Therefore, we attribute the larger variance for occurrence rate posteriors for the ``$planets\uparrow$'' cases in part to (a) the larger variance in the detection efficiency across this parameter space and (b) to the larger span in parameter space covered by the power law rate intensity parameterization. 
Furthermore, although we expect the ``$planets\uparrow$'' cases to have larger occurrence rates than the $planets\downarrow$ cases (because we are probing a larger domain) the ``$planets\uparrow$'' occurrence rate posteriors could be over estimated due to the low, unaccounted for, reliability in the corner near 0.75 $R_{\oplus}$ and 300 days.

\subsection{Sensitivity to Selected Stars}  \label{sec:results:senstostars}

Subtle differences in stellar cuts can impact the number of planets selected, where more (fewer) stars results in a smaller (larger) planet occurrence rate posterior mean.  
Table \ref{table4:StarPlanetCutNumbers} shows that the ``$DR25$ $GK \;cuts\, (stars) \downarrow$'' case has approximately $8\%$ fewer selected stars than the ``$DR25$ $GK \;cuts\, (stars) \uparrow$'' case.  
For these two selections of $GK$ star cuts used in this study, see Table \ref{table1:StarRanges}.
We compare occurrence rates across these subtle differences in selected stars to first assess how sensitive our occurrence rate posteriors are to the choice of target stars.
The difference in occurrence rates across subtle changes in stellar parameter cuts can be assessed by comparing the dashed-green/black and the dashed-blue/purple curve pairs in the top panel of Figure \ref{fig:f1} and the dashed-orange/pink and dashed-red/cyan curve pairs in the middle panel of Figure \ref{fig:f1}.
In these comparisons, the selected star parameters are varied while holding both the planet radius measurement uncertainty and the ranges in selected planet parameters fixed.

The ``$planets\downarrow$'' case has a larger difference ($8\%$) in the number of selected planets that make it through the two different $GK$ stellar cut designations than the ``$planets\uparrow$'', yet this has a smaller influence on the difference in occurrence rate modes between these stellar cut designations.
The ``$planets\uparrow$'' case has a smaller difference ($5\%$) in selected planets than the ``$planets\downarrow$'' case, and a larger difference in occurrence rate between occurrence rates calculated using these two stellar cut designations. 
The smaller difference in occurrence rate modes for the ``$planets\downarrow$'' cases is likely in part due to the smaller area in parameter space, which must be described by the power law rate intensity parameterization. 
Furthermore, differences in the occurrence rate posteriors between the two selections of $GK$ stars may be from differences in the signal to noise regime (e.g., the $GK \;cuts\; \downarrow$ regime containing slightly larger maximum stellar radii and slightly cooler stars may let through more false positives in the ``$planets\uparrow$'' case).
It's also possible that cooler $GK$ stars host more planets, because we see the slight increase in occurrence rate posterior means in both selections of planet parameters. 
\ct{Mulders2015} find that the occurrence of Earth to Neptune-sized planets is successively higher toward later spectral types at all orbital periods probed by $Kepler$.

\subsection{Sensitivity to planet radius measurement uncertainties} \label{sec:results:senstouncs}

Our analysis shows that when the planet-to-star radius ratio uncertainties are included, there is an upward shift in the occurrence rate posterior mean relative to when the planet-to-star radius ratio uncertainties are not included.   
In Figure \ref{fig:f1}, we can compare cases with fixed selected stars and planets for the DR25 catalogs, including measurement uncertainties in planet size (indicated by ``$\sigma_{R_{p}}\uparrow$'') and not including them (indicated by ``$\sigma_{R_{p}}\downarrow$'').
The  ``$\sigma_{R_{p}}\downarrow$''/``$\sigma_{R_{p}}\uparrow$'' pairs of marginal posteriors for the number of planets per $GK$ star are shown as dashed-green/dashed-blue and black/purple curve pairs in the top panel of Figure \ref{fig:f1}, and the dashed-orange/dashed-red and pink/cyan curve pairs in the middle panel of Figure \ref{fig:f1}, respectively.  

The upward shift in the occurrence rate posterior mean can largely be attributed to (a) the wide range in uncertainty values across the planet radius sample.  
For planets that have well constrained radius uncertainty, the location in completeness space stays relatively unchanged, whereas planets with large fractional radius uncertainties are more likely to have a large uncertainty in their detection completeness. 
(b) The detection probability is a sharp function of planet size near the detection threshold, with small planets more likely to be missed.  
For those small detected planets in the selected planets sample that have larger planet radius measurement uncertainties, their observed radius will be more biased relative to their true radius.  
Both of those effects cause the model that ignores uncertainties to be biased towards a lower occurrence rate for more selected planets with radii near the threshold of detection.

\subsection{Distribution Comparison to Burke+2015} \label{sec:results:pfmodel}

We use a joint power law rate intensity function in planet radius and orbital period for our inhomogeneous Poisson point process likelihood.
This generative model is specified to capture broad features of the results of planet \changes{formation and subsequent evolution} over small ranges in \changes{planet orbital period and radius.}
This likelihood and parameterization for \textit{Kepler} exoplanet occurrence rates was put forth in \ct{Youdin2011} and later applied by \ct{Burke2015}, but neither of these studies included measurement uncertainties in a hierarchical Bayesian statistical framework. 
We recreate the conditions of \ct{Burke2015} to benchmark our methods and to evaluate how occurrence rates have changed when using the latest $Kepler$ planet candidate catalog (the $DR25$ planet candidate catalog).
Our result for this occurrence rate is indicated as the dashed-grey curve in the middle panel of Figure \ref{fig:f1} and labeled ``$Q1-Q16$ $GK \;cuts\, \uparrow$'' for the ``$planets\uparrow$'' case, without measurement uncertainties (``$\sigma_{R_{p}}\downarrow$''), in the figure legend. 
In this case, we find an occurrence rate posterior mean of $0.85$ with a $68\%$ credible interval of $0.72$ to $0.99$, and an allowed range of 0.48 to 1.58. 
For the same set of stellar and planet parameter cuts, \ct{Burke2015} report an occurrence rate posterior mean of $0.77$ with an allowed range of 0.3 to 1.9. 
We attribute the smaller posterior width and larger posterior mean calculated in this study to be from a combination of unaccounted for differences in the custom catalog used in \ct{Burke2015} and the $Q1-Q16$ catalog available at the NASA Exoplanet Archive, and potentially due to differences in the MCMC methods and diagnostics used.

\subsection{Stars from $Gaia$}  \label{sec:results:starsGaia}

Using our statistical framework, we can compare disparate stellar catalogs. 
With the $Gaia$ updated stellar properties, the assumed stellar radii became larger on average \cp{Berger2018}.
Additionally, the sample now has fewer evolved stars for which $Kepler$ has reduced planet detection efficiency due to their larger size.
We first assess the impact of updated stellar radii from $Gaia$ on the mean and variance of occurrence rate posteriors in this region of parameter space. 
Updating stellar radii with $Gaia\,DR2$ parameter estimates will change both the precomputed completeness functions and the resulting planet sizes.
In this first case, we exclude the measurement uncertainties in planet size that come from both the uncertainty in $R_{p}/R_{*}$ measurements and from stellar radius measurements. 
Thus, we simply multiply the planet-to-star-radius ratios in the \textit{Kepler} $DR25$ catalog by the new stellar radii estimates from $Gaia\; DR2$, and change the stellar radius estimates used in the completeness model.
The resultant occurrence rate posterior is shown in the bottom panel of Figure \ref{fig:f1} as the brown curve labeled ``$DR25 + Gaia \;(GK \;cuts\; \downarrow)$'' with ``$planets\uparrow$'' cuts.
Comparing this occurrence rate posterior to the pink curve in the middle and bottom panels of Figure \ref{fig:f1} (``$DR25$ $GK \;cuts\, \downarrow$'' with ``$planets\uparrow$'' cuts) demonstrates the increase in the mean and variance of the occurrence rate posterior for the $Gaia$ updated planet radius point estimates and completeness inputs.  
The large difference in these two occurrence rates can be attributed to planets moving out of the planet radius range of interest, to changes in the precomputed completeness (due to shifting stellar radii values), and to planets and stars being removed from the sample when using more aggressive stellar cuts (described in \S \ref{sec:stars}).  

Next, we propagate the stellar radii uncertainties from $Gaia\; DR2$ into the planet size ($R_{p}$) uncertainties (and are now included along with the contribution to the planet radius uncertainty from $R_{p}/R_{*}$ measurements) while updating stellar radii in the precomputed completeness model. 
The resulting occurrence rate posterior is shown as the light-green curve in the bottom panel of Figure \ref{fig:f1}, exhibiting a much wider posterior (larger variance) than previous posteriors that did not include the contribution to the planet radius uncertainty due to the host star radii uncertainties. 
This marginal posterior has a $68\%$ credible interval of $0.49$ to $0.77$ and a mean of $0.63$. 
This occurrence rate posterior has a larger variance and a smaller posterior mean than the posterior for this parameter space using the $Q1-Q16$ planet candidate catalog, which has a $68\%$ credible interval of $0.72$ to $0.99$ and a posterior mean of $0.85$, and larger than the occurrence rate posterior when using the $DR25$ planet candidate catalog alone, which has a $68\%$ credible interval of $0.41$ to $0.59$ and a posterior mean of $0.50$. 
This shows that previous studies have overestimated the occurrence rate in this region of parameter space, likely because previous lower reliability \textit{Kepler} planet candidate catalogs, such as the $Q1-Q16$ catalog, likely included more false positives. 
However, selecting a cleaned stellar catalog partially compensates for this change.

\subsection{Results using Berger+2018 Catalog}  \label{sec:results:BergerCatalog}

\begin{figure}
\epsscale{1.0}
\plotone{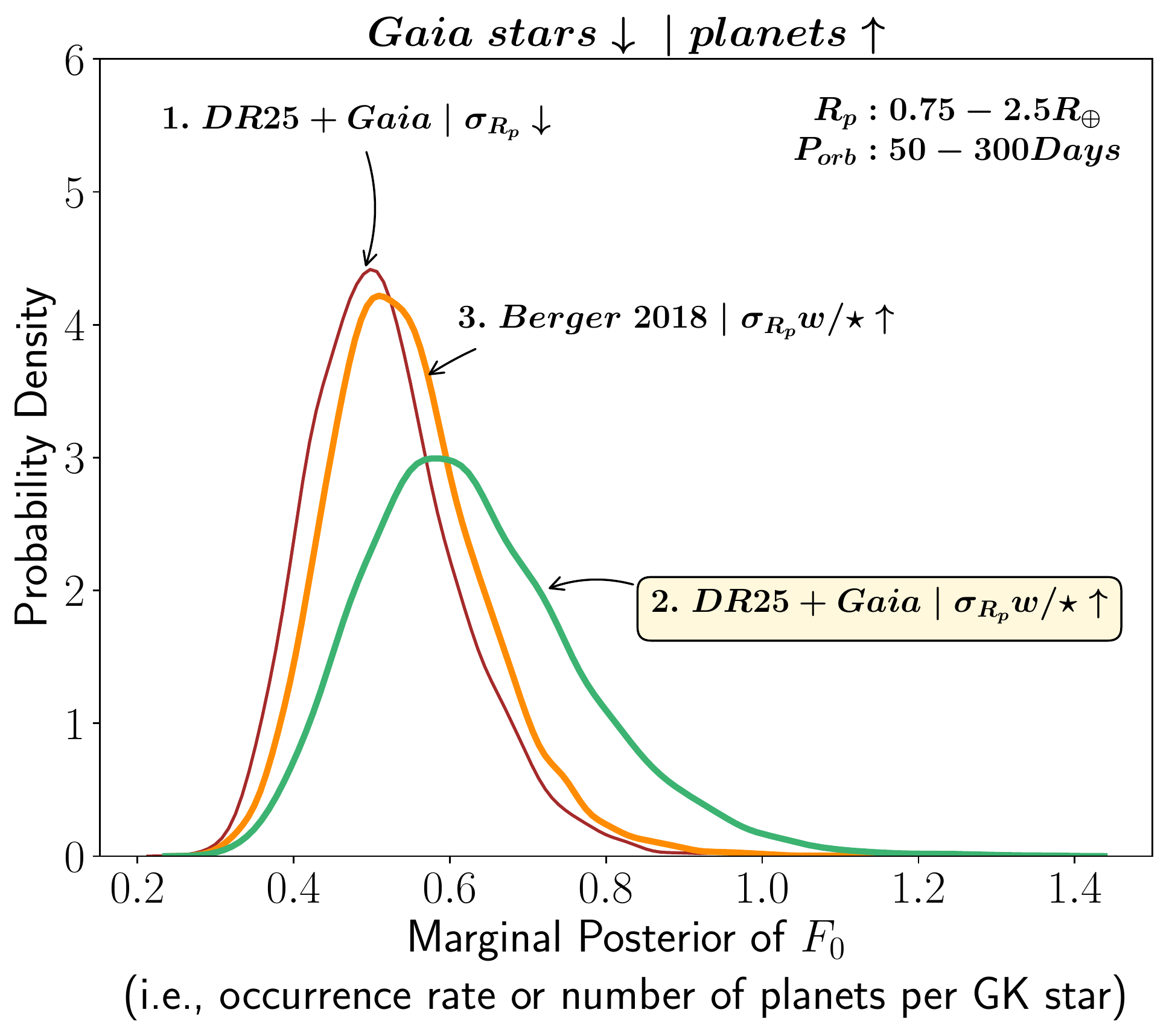}
\caption{\textbf{Kernel density estimates of marginal posterior distributions for the number of planets per GK star.} 
The posteriors shown here are for planets with radii between 0.75$R_{\oplus}$ and 2.5$R_{\oplus}$ and orbital periods between 50 and 300 days. 
The (1.) brown curve corresponds to the occurrence rate posterior calculated using the $DR25$ Kepler star and planet catalogs with stellar radii updated by crossmatching with $Gaia$ $DR2$ data, and does not include measurement uncertainty in planet size. 
The (2.) green curve is this same case, but now includes planet radius measurements uncertainties from the $R_{p}/R_{*}$ measurements and from the uncertainties in stellar radii measurements when using $Gaia$ data.
The (3.) orange curve is the occurrence rate posterior when using the \ct{Berger2018} catalog that includes stellar parameters updated using spectroscopic followup for host stars only. 
The stellar sample in full is updated using $Gaia$ $DR2$.  
The orange curve demonstrates that using heterogeneous stellar parameters introduces a large bias in occurrence rates.
\label{fig:f2}}
\end{figure}

The \ct{Berger2018} study has provided a catalog of revised planet and star radius measurements using $Kepler$ $DR 25$ stars crossmatched with stars from $Gaia\;DR2$.
\ct{Berger2018} use quality cuts similar to those described in \S \ref{sec:stars} and also incorporate the stellar host star spectroscopic followup from the \textit{California-Kepler Survey} (CKS) \cp{Fulton2017, Petigura2017, Johnson2017}.
Furthermore, the results from \ct{Berger2018} account for the impact of reddening.
The orange curve in Figure \ref{fig:f2} shows the marginal posteriors for the number of planets-per-$GK$ star, over the $stars\downarrow$ and $planets\uparrow$ parameter space, using the \ct{Berger2018} catalog. 
For this case, we find a $68\%$ credible interval of $0.45$ to $0.64$ and a posterior mean of $0.55$.
This result is close to the result for the occurrence rate posterior distribution using the $DR25$ + $Gaia$ crossmatch (without updates using CKS) described in \S \ref{sec:stars}, when measurement uncertainties are not included (shown as the brown curve in Figures \ref{fig:f1} and \ref{fig:f2}). 
The results when measurement uncertainties are included for the $DR25$ + $Gaia$ catalogs (without spectroscopic host star followup) is shown as the green curve in Figures \ref{fig:f1} and \ref{fig:f2}, for reference. 
Systematic differences in measurement uncertainties for stars with and without detected planets are not included in our statistical model. 
\changes{The orange curve shown in Figure \ref{fig:f2} demonstrates that using the \ct{Berger2018} catalog that includes heterogeneous stellar parameters introduces a bias in the occurrence rate. We note that we have not included the impact of reddening in our catalog, which could impart differences in the base occurrence rate calculated before planet radii uncertainties are included. However, this would not account for the bias we see between the orange and green curves were planet radius uncertainties are included in the model.} \changes{We note that this bias can be addressed with spectroscopic characterization of all stars in the catalog.}

\begin{deluxetable*}{llccccr}
\tablecaption{Summary Statistics for Occurrence Rate Posterior Distributions}
\tablenum{5}

\tablehead{\colhead{} & \colhead{distribution} & \colhead{mean} & \colhead{var} & \colhead{std} & \colhead{mode} & \colhead{68\% Credible Interval} }
\startdata
$\boldsymbol{planets\downarrow}$&$\boldsymbol{DR25\;|\;  \sigma_{R_{p}}\downarrow\;|\;  stars\uparrow}$ & $ 0.154 $ & $ 0.001 $ & $ 0.029 $ & $ 0.145 $ & $ [0.126, 0.182] $  \\ 
&$\boldsymbol{DR25\;|\; \sigma_{R_{p}}\downarrow\;|\; stars\downarrow}$ & $ 0.168 $ & $ 0.001 $ & $ 0.03 $ & $ 0.157 $ & $ [0.139, 0.198] $  \\ 
&$\boldsymbol{DR25\;|\;  \sigma_{R_{p}}\uparrow\;|\;  stars\uparrow}$ & $ 0.188 $ & $ 0.002 $ & $ 0.04 $ & $ 0.174 $ & $ [0.149, 0.226] $  \\ 
&$\boldsymbol{DR25\;|\; \sigma_{R_{p}}\uparrow\;|\;  stars\downarrow}$ & $ 0.199 $ & $ 0.002 $ & $ 0.04 $ & $ 0.179 $ & $ [0.16, 0.239] $  \\ 
\hline
$\boldsymbol{planets\uparrow}$&$\boldsymbol{DR25\;|\; \sigma_{R_{p}}\downarrow\;|\;  stars\uparrow}$ & $ 0.35 $ & $ 0.003 $ & $ 0.055 $ & $ 0.322 $ & $ [0.297, 0.405] $  \\ 
&$\boldsymbol{DR25\;|\;  \sigma_{R_{p}}\downarrow\;|\;  stars\downarrow}$ & $ 0.407 $ & $ 0.004 $ & $ 0.062 $ & $ 0.385 $ & $ [0.348, 0.467] $  \\ 
&$\boldsymbol{DR25 \;|\; \sigma_{R_{p}}\uparrow \;|\; stars\uparrow}$ & $ 0.442 $ & $ 0.007 $ & $ 0.085 $ & $ 0.4 $ & $ [0.361, 0.523] $  \\ 
&$\boldsymbol{DR25\;|\; \sigma_{R_{p}}\uparrow\;|\;  stars\downarrow}$ & $ 0.497 $ & $ 0.008 $ & $ 0.089 $ & $ 0.452 $ & $ [0.411, 0.585] $  \\ 
&$\boldsymbol{Q1-Q16\;|\;  \sigma_{R_{p}}\downarrow\;|\;  stars\uparrow}$ & $ 0.854 $ & $ 0.017 $ & $ 0.131 $ & $ 0.812 $ & $ [0.724, 0.985] $  \\ 
&$\boldsymbol{DR25+Gaia\;|\;  \sigma_{R_{p}}\downarrow\;|\;  Gaia \;stars\downarrow}$ & $ 0.519 $ & $ 0.01 $ & $ 0.098 $ & $ 0.483 $ & $ [0.424, 0.613] $  \\ 
&$\boldsymbol{DR25+Gaia\;|\;  \sigma_{R_{p}} w/\star\uparrow\;|\;  Gaia \;stars\downarrow}$ & $ 0.63 $ & $ 0.02 $ & $ 0.141 $ & $ 0.606 $ & $ [0.492, 0.767] $  \\ 
\enddata

\tablecomments{We summarize the occurrence rate posterior distributions from Figures \ref{fig:f1} and \ref{fig:f2} via the mean, variance, standard deviation, mode, and the 68.3\% credible interval. The credible intervals are calculated such that the left and right hand regions of the posterior distribution outside the credible interval are equal in area.}

\label{table5:SumStats}
\end{deluxetable*}

\section{Discussion}  \label{sec:discussion}

The application of hierarchical Bayesian inference to infer planet occurrence rates handles a relatively small number of detected planets by pooling and mustering the strength of each constituent while learning about the population. 
By using Hamiltonian Monte Carlo to sample from our posterior, we can apply a high-dimensional hierarchical Bayesian model that has more parameters than measurements.  
\changes{There are more parameters than measurements because when we include planet radius measurement uncertainties (i.e., noise, error, and model assumptions), the planet radius for every planet in the model becomes a parameter in the model. }
By assessing how the occurrence rates behave in response to subtle difference in the inputs, we can see the positive impact of the \textit{Kepler} science team's efforts to provide high quality occurrence rate data products, and we can evaluate the opportunities for advancing the depth of the science questions we are asking regarding exoplanetary systems.  

Current analysis from $Gaia$ data has provided stellar radii with average uncertainties of 8\% \citep{Berger2018}.
Our selected $Gaia$ crossmatched stellar population has uncertainties of approximately $5\%$ on average.  
This allows us to incorporate quality stellar data into the current occurrence rate framework we are using, parameterized by planet orbital period and planet radius. 

Hierarchical parametric Bayesian exoplanet occurrence rate studies provide the foundation for constraining more complex exoplanet population distributions using \textit{Kepler} data. 
As the data quality improves with complementary observations such as stellar follow-up, and with reprocessing of the current \textit{Kepler} data using emerging statistical methods, scientists can begin to answer more in-depth questions in order to characterize planetary systems.  
In the following section we discuss occurrence rates from several angles: the population model, the data quality, the computational methods used to constrain hierarchical Bayesian models, and the science questions at hand.

\subsection{Generative model and precomputing the survey completeness} \label{disc:precompcomp}

The likelihood we use in this study assumes a rate intensity that is correlated between bins, similar to \ct{Burke2015} and \ct{ForemanMackey2014}.
This is important to consider when including planet radius measurement uncertainty in occurrence rate studies, since each planet's size can now take on a variety of values.  
In this case, the data generating process would be the outcome of planet \changes{formation and subsequent evolution}, whereas a non-parametric Bayesian method such as a Gaussian Cox Process would be agnostic to any planet \changes{formation and subsequent evolution} relations.   

In this initial study, we use a precomputed completeness grid over planet radius and orbital period described in \S \ref{obs:detmodel}.
When assessing the impact on occurrence rates from planet radius measurement uncertainties, our precomputed completeness grid eases the computation. 
In order to include the contribution from the host star radius into the planet radius uncertainty, we need to include the host star uncertainty into the calculation of the probability of detection, the geometric transit probability, and any functions in the completeness model that depend on stellar properties. 
In \S \ref{sec:results:starsGaia}, we probe how occurrence rate posteriors change when using stellar properties from $Gaia$ to update the stellar radius point estimates for each observed star, the means of the planet candidate radii measurement uncertainties, and the means of the host star radii measurement uncertainties.
In this case we assess the impact of the contribution to the planet radius uncertainty from the host star radius uncertainty by approximating the completeness function as constant within each bin in planet radius and orbital period.  
We find occurrence rate marginal posterior distributions are not changed when increasing the resolution of our completeness grid.

\subsection{Future work} \label{sec:ongoingwork}

Future studies to include the stellar radii uncertainties into the completeness model and therefore include the stellar radii as latent variables in our hierarchical Bayesian model, may require the calculation of the completeness model in each iteration when sampling from the likelihood.
This would replace the precomputed completeness we use in this study, which is used as input in our statistical framework.   
The analytic completeness models described in \ct{Burke2015} and used in this work, take significantly less computational time than the numerical completeness functions available as part of the $DR25$ $Kepler$ occurrence rate data products.  
Moving away from a precomputed completeness and using the latest numerical completeness models may require more advanced computing resources and techniques to constrain occurrence rate statistical frameworks that include stellar parameter measurement uncertainties.    

Measurement uncertainties for orbital periods are negligible, but when re-parameterizing in terms of insolation flux, uncertainties in stellar effective temperature, stellar multiplicity, stellar mass and stellar radius could contribute significantly to the uncertainties in occurrence rates as a function of insolation flux.  
Updates to stellar effective temperatures from analysis of $Gaia$ data will allow future studies to properly parameterize the occurrence rate in terms insolation flux, as orbital distance is calculated from the stellar mass and orbital period, and the orbital distance estimate is used in the detection efficiency calculations. 
By including the completeness functions directly into the hierarchical Bayesian model's data generating process (instead of a precomputed completeness) in addition to a functional form for the planet \changes{formation and subsequent evolution} model, it will be possible to marginalize over uncertainties in stellar parameters.
This will ultimately lead to constraining occurrence rates as a function of insolation flux (and other stellar parameters) in addition to planet parameters.

By using $Gaia$ data to better constrain planet radius uncertainties and provide accurate fractional uncertainties for insolation flux, we can assess the impact of excluding measurement uncertainty in the occurrence rate parameterizations that go beyond the impact of the planet radii uncertainties investigated here.
This will improve previous occurrence rates calculated in terms of insolation flux that are biased by the inverse detection efficiency method (e.g., \ct{Fulton2018}). 
The large disparity in the number of selected stars for the different catalogs used to investigate changes in occurrence rate posteriors motivates including stellar parameter dependence directly into occurrence rate studies in the future.

When crossmatching the $DR25$ \textit{Kepler} stellar catalog with the $Gaia\;DR2$ stellar parameters, we remove stars that have indications they may be poorly-resolved binaries.
This provides results that are less contaminated with dilution from binarity than previous studies.
\ct{Ciardi2015, Hirsch2017, Furlan2017a, Furlan2017b, Furlan2018} have measured a non-negligible planet radius correction factor to account for stellar multiplicity.
Furthermore, \ct{Bouma2018} show that for terrestrial-sized planets, stellar multiplicity can contribute uncertainties in occurrence rates of approximately 50\%.
Stellar multiplicity is an important consideration for occurrence rates beyond the dilution of the planet radius by over estimating the size of its host star, as it can also impact the measured semi-major axis. 
In future studies, including a model of the impact of stellar binarity directly into the generative model used in this analysis will allow the impact of stellar binarity on occurrence rates to be measured.  

Preliminary occurrence rate estimates of potentially habitable planets are lower with the new reliability estimates from the $DR25$ 9.3 \textit{Kepler} occurrence rate data products. 
This suggests that a vigorous treatment of the catalog reliability for occurrence rate studies will be necessary for learning about the population of potentially habitable planets. 

By including planet radius measurement uncertainties into a parametric hierarchical Bayesian occurrence rate calculation, we have provided the foundation for researchers to use the \textit{Kepler} dataset to constrain parameters in analytic planet distribution models.
This can be done by investigating these relations in place of the simplistic power law intensity parameterization described in this work. 
Furthermore, \ct{Zink2019} show that the Kepler dichotomy can be filled in by accounting for the effects of multiplicity on the detection efficiency, and provide improved estimates of the multiplicity distribution.  
Future studies can include this updated detection efficiency while also incorporating radius measurement uncertainties into the likelihood function.

\section{Conclusion}  \label{sec:conclusions}

When using our parametric hierarchical Bayesian model in conjunction with $Gaia$ data to (i) remove stars that have indications they may be poorly-resolved binaries; (ii) update the uncertainties in planet radii and in turn include the contribution of the host star radii into the uncertainty in planet radii; and (iii) update the stellar parameters in the completeness model, 
\begin{itemize} 
\item we estimate the $GK$ star planet occurrence rate between 0.75 and 2.5 $R_{\oplus}$ and 50 to 300 days to have a $68\%$ credible interval of $0.49$ to $0.77$ and a mean of $0.63$.
\end{itemize}
When using the \ct{Berger2018} catalog that includes spectroscopic followup of host stars only, Gaia updated stellar radii, and reddening,
\begin{itemize}
\item \changes{we find that a bias is introduced into the occurrence rate posterior distributions when using heterogeneous stellar radii measurement uncertainties.} 
\end{itemize}

By performing a hierarchical Bayesian occurrence rate analysis in a particular part of planet parameter space with differences in reliability and completeness, 
\begin{itemize}
\item we find an upward shift in the occurrence rate posterior mean and a larger posterior variance when including measurement uncertainty in planet radius.
\end{itemize}

When evaluating the sensitivity of planet occurrence rates to subtle changes in the selected stars, 
\begin{itemize}
\item our results suggest that our hierarchical Bayesian models (Bayesian models that include measurement uncertainties) are less sensitive to subtle differences in stellar properties, and more so to the the selected ranges in planet parameters.
\end{itemize}
By evaluating a set of slightly cooler stars and a set of slightly warmer stars across a two sets of selected planets with different completeness and reliability characteristics \begin{itemize}
\item we show that the choice of stellar cuts can influence the number of planet candidates selected over the planet radius and orbital period grid of interest.  
\item we find that the cooler star sample has a slightly higher occurrence rate posterior for both sets of selected planets. 
\end{itemize} 
This difference could in part be from (a) the slightly cooler selected stars letting through more false positives, and (b) the slightly cooler set of $GK$ stars could host more planets. 
Work by \ct{Dressing2013, Dressing2015} found cooler M Dwarfs stars have larger occurrence rates. \changes{Note that the \ct{Dressing2013, Dressing2015} results are derived for much cooler stars 2600-4000K than those investigated here.}
This motivates the inclusion of a more vigorous treatment of the catalog reliability in future occurrence rate studies. 
Furthermore, the inclusion of stellar population level parameters in hierarchical Bayesian occurrence rate studies will allow the characterization of the stellar dependence of exoplanet occurrence rates. 
It may be important to include the stellar dependence in statistically robust occurrence rate studies before we can select targets of opportunity for some exoplanet research. 

We also evaluate the impact of selecting planets in a slightly higher average completeness space, compared to a part of parameter space with slightly less average and larger variance in completeness.
\begin{itemize}
\item We find that the selection of planets over the slightly more complete part of parameter space results in occurrence rate marginal posteriors with less variance than the space evaluated over a slightly less complete part of parameter space with more variance in completeness.
\end{itemize} 
This is interesting because the ``$planets\downarrow \,$'' case (slightly more complete space) contains approximately $50\%$ less planets than the ``$planets\uparrow$'' case.
\begin{itemize}
\item This suggests that the precision (variance) in the occurrence rate posteriors when using the statistical framework in this work is less sensitive to the number of planets that make it through the planet cuts, and more so on the (a) span that the rate intensity parameterization is providing coverage over and (b) the effective number of stars searched.
The effective number of stars searched (i.e., how efficient \textit{Kepler} is at detecting planets in a given part of parameter space) depends on the characteristics of the completeness and reliability space, and the signal-to-noise regime. 
\end{itemize}

\acknowledgments
Shabram's research is supported by an appointment to the NASA Postdoctoral Program at the NASA Ames Research Center, administered by Universities Space Research Association under contract with NASA.
This manuscript benefited from discussions at the Stanford Stan User's Group. 
This work benefited from Dan Foreman-Mackey's blog post regarding an experiment in open science for \textit{Kepler} exoplanet occurrence rates.  
This work makes use of the Stan open source Bayesian modeling language, Jupyter, and Python software including the SciPy ecopsystem and Seaborn.
E.B.F. and D.C.H. acknowledge the Penn State Center for Astrostatistics and Center for Exoplanets and Habitable Worlds, which is supported by the Pennsylvania State University, the Eberly College of Science, and the Pennsylvania Space Grant Consortium. Computations for this research were performed on the Pennsylvania State University's Institute for CyberScience Advanced CyberInfrastructure (ICS-ACI).  E.B.F. acknowledges support from NASA Exoplanets Research Program award \#NNX15AE21G. This work was partially supported by the NSF grant DMS 1127914 to the Statistical and Applied Mathematical Sciences Institute (SAMSI).  
The results reported herein benefitted from collaborations and/or information exchange within NASA's Nexus for Exoplanet System Science (NExSS) research coordination network sponsored by NASA's Science Mission Directorate.
T.A.B and D.H. acknowledge support by the National Science Foundation (AST-1717000) and the National Aeronautics and Space Administration under Grants NNX14AB92G and NNX16AH45G

\software{Stan \cp{StanPaper}, SciPy\cp{SciPyEcosystem}, Seaborn\cp{Seaborn}, Jupyter \cp{Kluyver:2016aa}}

\begin{appendix}
\section{Two sample Kolmogorov-Smirnov statistics for occurrence rate posterior distribution comparisons}

\counterwithin{figure}{section}
\setcounter{figure}{0}
In Figure \ref{fig:fA1}, we show the ``two sample Kolmogorov-Smirnov (K-S) statistic'' to asses the distance between pairs of occurrence rate posterior distributions. A K-S statistic close to 0 means the distributions are likely both drawn from the same underlying population and a K-S statistic of 1 means it is less likely the distributions come from the same underlying distribution.  The label colors correspond to the distribution plot color in Figures \ref{fig:f1} and \ref{fig:f2}.

\begin{figure}
\epsscale{1.0}
\plotone{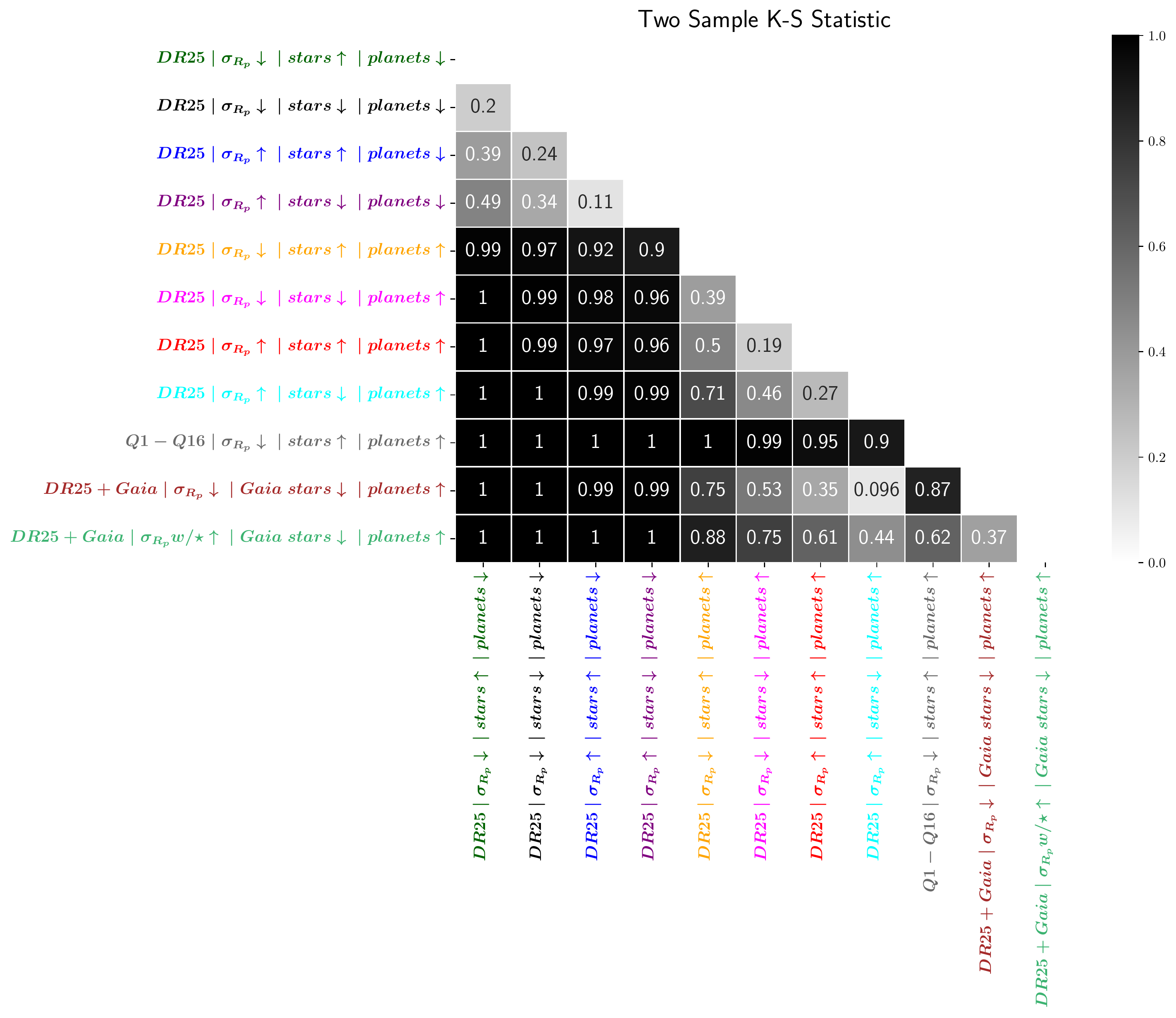}
\caption{\textbf{Two sample Kolmogorov-Smirnov statistics for occurrence rate posterior distribution comparisons} } \label{fig:fA1}
\end{figure}

\end{appendix}

\clearpage
\bibliographystyle{aasjournal}



\end{document}